\documentstyle[twoside,fleqn,npb,epsfig,amsmath]{article}

\newcommand{\abbrev}{\sf\small}

\newcommand{\fig}[1]{Fig.\,\ref{#1}}

\newcommand{\sct}[1]{Sect.\,\ref{#1}}

\newcommand{\lo}{{\abbrev LO}}
\newcommand{\nlo}{{\abbrev NLO}}
\newcommand{\nnlo}{{\abbrev NNLO}}
\newcommand{\mhiggs}{M_H}
\newcommand{\reference}[1]{Ref.\,\cite{#1}}

\newcommand{\Li}{{\rm Li}}
  \title{Scalar and Pseudo-Scalar Higgs Production at Hadron Colliders}
  
  \author{Robert Harlander\address{TH Division,
      CERN, CH-1211 Geneva 23, Switzerland}
    and
    William Kilgore\address{HET, Physics Department, Brookhaven
      National Laboratory, Upton, NY 11973, USA}}

\begin{document}

%- }}}
%- {{{ abstract:

\begin{abstract}
The evaluation of the {\abbrev NNLO} {\abbrev QCD} corrections
to the production of a scalar and a pseudo-scalar Higgs boson is
described.
\end{abstract}

%- }}}
%- {{{ coverpage:

\def\thefootnote{\fnsymbol{footnote}}

\mbox{}
\vspace{-10em}
\begin{flushright}
{\tt BNL-HET-02/23, CERN-TH/2002-326, hep-ph/0211380}\\
\end{flushright}
\vspace{4em}

\mbox{}\\
{\Large Scalar and Pseudo-Scalar Higgs Production at Hadron
  Colliders\footnote{ Talk given by R.H.\ at the ``6th International
    Symposium on Radiative Corrections (RADCOR 2002) /
    Loops and Legs in Quantum Field Theory'', September 8--13, 2002,
    Kloster Banz, Germany.}
}\\[1em]
Robert Harlander$^a$ and William Kilgore$^b$\\[1em]
$^a${\rm TH Division,
    CERN, CH-1211 Geneva 23, Switzerland}\\[.5em]
$^b${\rm HET, Physics Department, Brookhaven
      National Laboratory, Upton, NY 11973, USA}\\[2em]

{\small The evaluation of the {\abbrev NNLO} {\abbrev QCD} corrections
  to the production of a scalar and a pseudo-scalar Higgs boson is
  described.}

\newpage
\mbox{}
\newpage

\setcounter{footnote}{0}
\setcounter{page}{1}
\def\thefootnote{\arabic{footnote}}

%- }}}
%- {{{ Introduction:

\maketitle

\section{Introduction and Motivation}
The experimental discovery of a Higgs boson could to be one of the
next big milestones in particle physics. If a Higgs boson exists, it
will most certainly be found at the Large Hadron Collider ({\abbrev
  LHC}), or even before, at the Tevatron.

The Higgs sector is the least explored part of the Standard Model. In
particular, it is unclear if really the minimal model with a single
Higgs doublet is realized in Nature.  Extended models, like the Minimal
Supersymmetric Standard Model ({\abbrev MSSM}), predict a larger variety
of Higgs bosons which differ among each other for example by their mass,
charge, CP-parity, and couplings.

Hints towards physics beyond the Standard Model could be obtained from
measuring these properties of a Higgs boson, once discovered. In this
respect, it is essential to understand the theoretical values of these
quantities in the framework of the Standard Model as precisely as
possible.

A clear understanding of the Higgs properties is often based on the
precise knowledge of the production rates. One can distinguish four main
production mechanisms at hadron colliders: Gluon fusion, associated
production with weak bosons, weak boson fusion, and
associated production of the Higgs boson with a top--anti-top pair. The
theoretical progress in each one of these modes has been enormous over
the past few years, both for the signal and the background processes
(see, e.g., \reference{CarHab} for a review and the corresponding
references). In the following we will only focus on the signal cross
section in the gluon fusion channel.

%- }}}
%- {{{ Gluon fusion:

\section{Gluon fusion}
\begin{figure}
  \begin{center}
    \leavevmode
    \begin{tabular}{c}
      \epsfxsize=10em
      \epsffile[145 450 460 660]{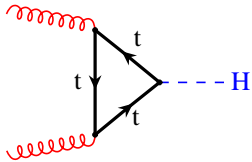}\\[-1em]
    \end{tabular}
    \caption[]{\label{fig::lo}\sloppy
      Leading order diagram for Higgs production in gluon fusion.
      }
  \end{center}
\end{figure}

The gluon fusion mode has been shown to be the dominant production
mechanism for Higgs bosons at hadron colliders a long time
ago~\cite{lo}.  The coupling of the Higgs boson to gluons is mediated by
a top quark triangle, so that the leading order process is described by
the one-loop diagram shown in \fig{fig::lo}. 

The fact that the \lo{} process already requires a one-loop calculation
makes the evaluation of higher orders even more challenging. It turns
out, however, that there is an approximation to the problem that
simplifies the task of evaluating higher orders significantly, without
much loss in theoretical accuracy. The relevant limit is given by taking
the top quark very heavy as compared to the Higgs boson.  Keeping the
full top mass dependence at \lo{} and evaluating the higher orders in
the heavy top limit leads to a very good approximation to the exact
result. This was explicitly demonstrated at \nlo{}~\cite{nlo},
where the full top mass dependence is known.

In the heavy top limit, the original diagrams factorize into a massive
component with vanishing external momenta, and a massless component with
the physical momenta of the in- and outgoing particles.  The massive
component represents an ``effective coupling constant'' $C(\alpha_s)$
which multiplies the $ggH$ interaction vertex. $C(\alpha_s)$ can be
evaluated perturbatively and is known up to N$^3$LO~\cite{coef}.

%- }}}
%- {{{ NNLO calculation:

\section{NNLO calculation}\label{sec::nnlo}
The sub-processes to be evaluated for the \nnlo{} rate are
\begin{itemize}
\item $gg\to H$ up to two-loop level
\item $gg\to Hg$, $qg\to Hq$, $q\bar q\to Hg$ up to one-loop level
\item $gg\to Hgg$, $gg\to Hq\bar q$, $qg\to Hqg$, $q\bar q\to Hq\bar q$,
  $q\bar q\to Hgg$, $qq\to Hqq$ at tree level
\end{itemize}
All the one-loop processes can be evaluated analytically using known
loop and phase space integrals. On the other hand, the evaluation of the
two-loop virtual diagrams as well as the phase space integration of the
double real emission contributions required new techniques, even though
a very similar process, the Drell-Yan production of lepton pairs, had
been evaluated up to \nnlo{} about ten years before~\cite{DY}.

The initial impulse was provided in a paper by Baikov and
Smirnov~\cite{baismi}.\footnote{R.H.\ is indebted to A.~Czarnecki for
  initially pointing out the relevance of this paper.}  It contains a
recipe to evaluate the three-point functions at two loops in complete
analogy to massless three-loop propagator diagrams. The evaluation of
the latter is standard in the field of multi-loop calculations. It can
be done with the help of the {\tt FORM} program {\tt
  MINCER}~\cite{form}.  Using \reference{baismi}, one can evaluate
general massless two-loop vertex diagrams with two massless legs in a
similar fashion. The corresponding integration routine can easily be
connected to the diagram generator {\tt QGRAF}~\cite{qgraf} in the
framework of the package {\tt GEFICOM}~\cite{geficom} (see also
\reference{auto}), so that the evaluation of the \nnlo{} virtual
contributions to $gg\to H$ is done fully automatically~\cite{virtual}.

Let us now describe the evaluation of the phase space integrals for the
emission of two gluons or quarks. In a first step, they were evaluated
in the soft limit, i.e., when the momenta of the final state gluons (or
quarks) tend to zero. This limit suffices to cancel the infra-red
divergences of the virtual and soft single-emission contributions, so
that after mass factorization one obtains a finite result.  The result
was then combined with the previously known collinear terms $\propto
\ln^3(1-x)$~\cite{kls} ($x \equiv\mhiggs^2/\hat s$) to arrive at the
so-called ``soft+sl''~\cite{soft} or ``{\abbrev SVC}''~\cite{cfg}
approximation.  Numerical differences between the final results of
\reference{soft} and \cite{cfg} could be attributed partly to the
different treatment of formally subleading terms $\propto \ln^n(1-x)$,
$n=0,1,2$. The remaining difference was due to the different sets of
parton distribution functions: No published set of \nnlo{} parton
distributions was available at that time, so that \reference{soft} used
{\abbrev NLO} sets for the \nnlo{} curves, while \reference{cfg} used
unpublished \nnlo{} sets.

The soft limit was used as a starting point for the method to evaluate
the complete \nnlo{} result~\cite{HKggh}. This means that a systematic
expansion of the partonic cross section around the soft limit $x \to 1$
was constructed, where $\hat s$ is the partonic c.m.s.\ energy. This
leads to a series in $(1-x)^n$, whose coefficients depend
logarithmically on $(1-x)$, up to $\ln^3(1-x)$ at \nnlo{}.  The series
was evaluated analytically up to $n=16$~\cite{HKggh}. The resulting
hadronic cross section was obtained from this by convolution with the
proper parton distributions. For $n>5$, its prediction is almost
independent on $n$, indicating that small values of $x$ have no
significant influence on the final result. The physical hadronic rate is
thus perfectly described in this approach.

The series in $(1-x)$ as described above, taken up to $n=\infty$, is
nothing but a representation of a sum of tri-, di-, and simple
logarithms,\footnote{This can be deduced from the known \nnlo{} result
  for the Drell-Yan process~\cite{DY,HKggh}.} multiplied by powers of of
$x$, $1/x$, and $1/(1-x)$, just like, e.g.,
\begin{equation}
\begin{split}
  \frac{\pi^2}{6} -
  \sum_{n=1}^{\infty}\frac{(1-x)^n}{n}\left(\frac{1}{n} -
    \ln(1-x)\right)
\end{split}
\end{equation}
can be viewed as a representation of $\Li_2(x)$.  Thus, making an ansatz
for the resummed result with unknown coefficients for these functions
and expanding it in terms of $(1-x)$, one can actually determine the
coefficients from the {\it truncated} series, by solving a system of
linear equations.  This requires the knowledge of a sufficient number of
terms in the expansion, of the order of 100 in our case.  The required
efforts for the evaluation and manipulation of the intermediate
expressions are quite remarkable.  The resummation has been achieved in
\reference{kichep02}, and the result is identical to the expression of
\reference{AnaMel}, obtained through a completely different method (see
also \reference{AnaRADCOR}).  For the final numerical results presented
below, however, it is irrelevant if we use the truncated series of
\reference{HKggh}, or the closed form of \reference{AnaMel}.

In order to arrive at a consistent \nnlo{} result, it is not sufficient
to evaluate the partonic cross section up to \nnlo{}. One also needs to
account for the parton evolution up to the same order.  So far, the
exact evolution kernels are not known. Thus, until the exact \nnlo{}
evolution becomes available, we use the approximate \nnlo{} parton set
of \reference{mrstnnlo} which is based on moments of the structure
functions~\cite{moments}.

%- }}}
%- {{{ Results:

\section{Results}

%- {{{ fig::lo:

%- }}}
%- {{{ fig::snnlo:

\begin{figure*}
  \begin{center}
    \leavevmode
    \begin{tabular}{cc}
      \epsfxsize=17em
      \epsffile[110 265 465 560]{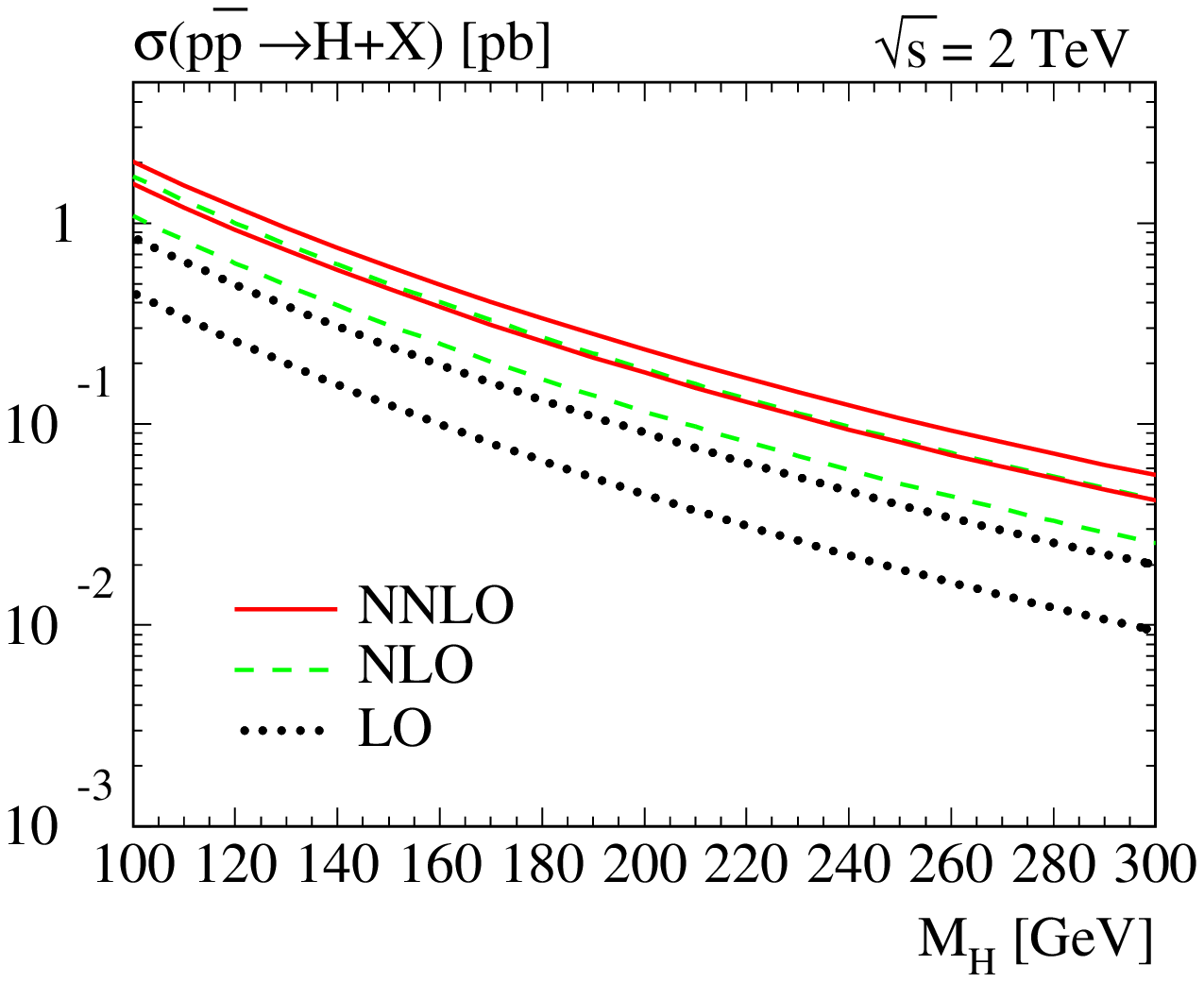}
      &
      \epsfxsize=17em
      \epsffile[110 265 465 560]{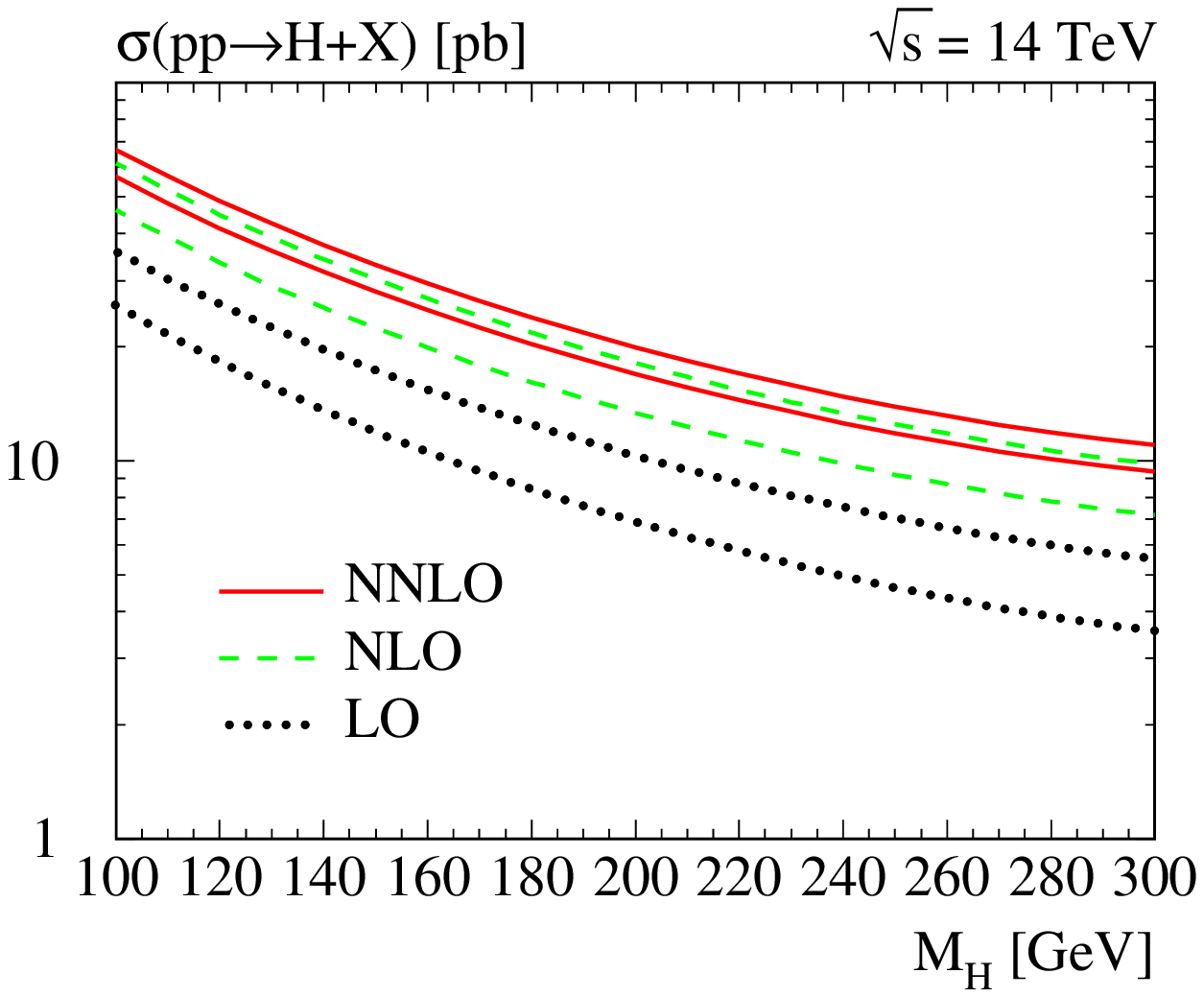}\\[-1em]
      $(a)$ & $(b)$
    \end{tabular}
    \caption[]{\label{fig::snnlo}\sloppy Total cross section for Higgs
      production in gluon fusion at $(a)$ Tevatron Run~II, and $(b)$ the LHC
      at leading (dotted), next-to-leading (dashed), and next-to-next-to
      leading order (solid). The upper (lower) curve of each pair
      corresponds to a choice of the renormalization and factorization
      scale of $\mu_R=\mu_F=M_H/2$ ($\mu_R=\mu_F=2\,M_H$).  }
  \end{center}
\end{figure*}

%- }}}
%- {{{ fig::scale2:

\begin{figure*}
  \begin{center}
    \leavevmode
    \begin{tabular}{ccc}
      \epsfxsize=12em
      \epsffile[184 210 411 610]{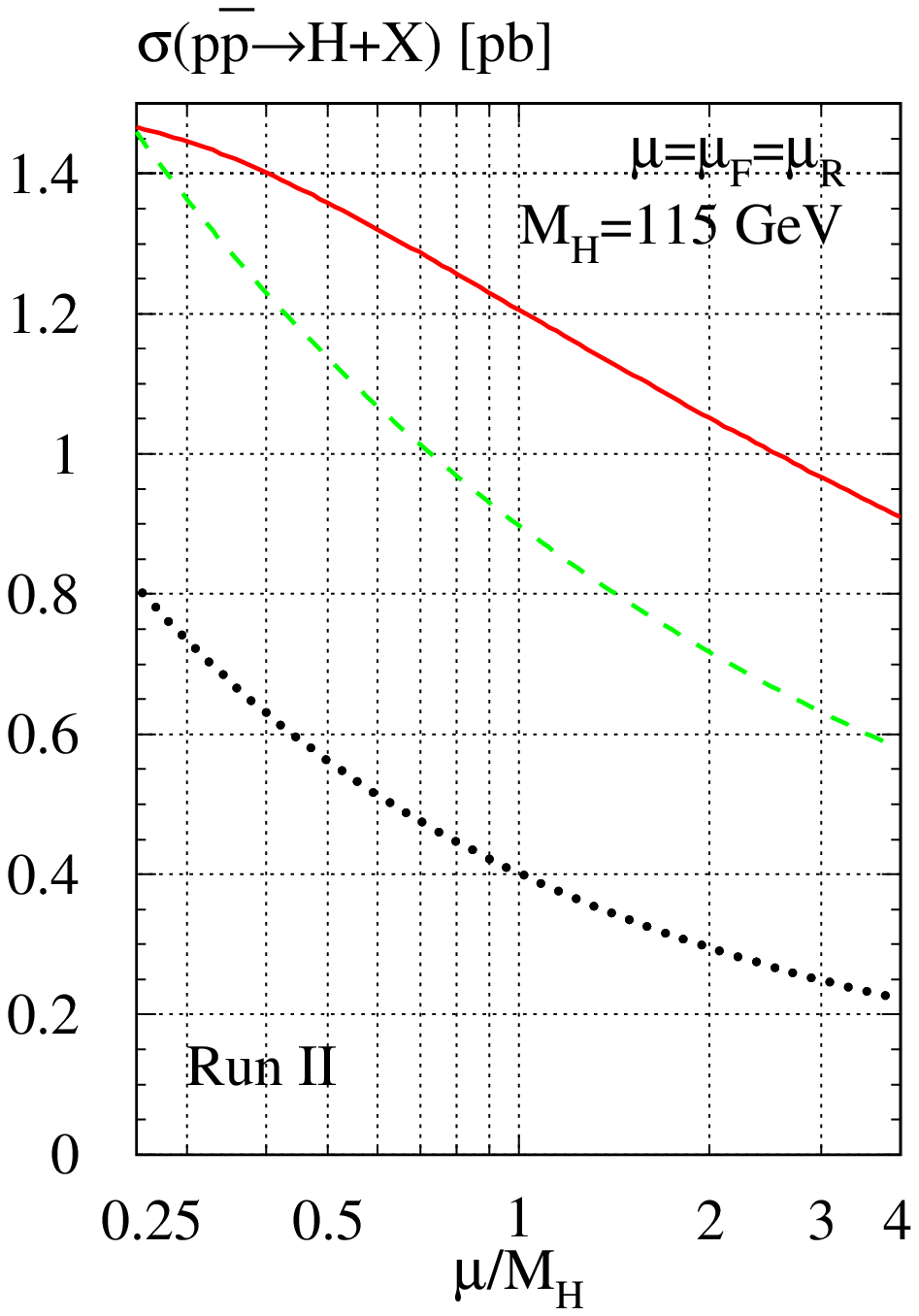} &
      \epsfxsize=12em
      \epsffile[184 210 411 610]{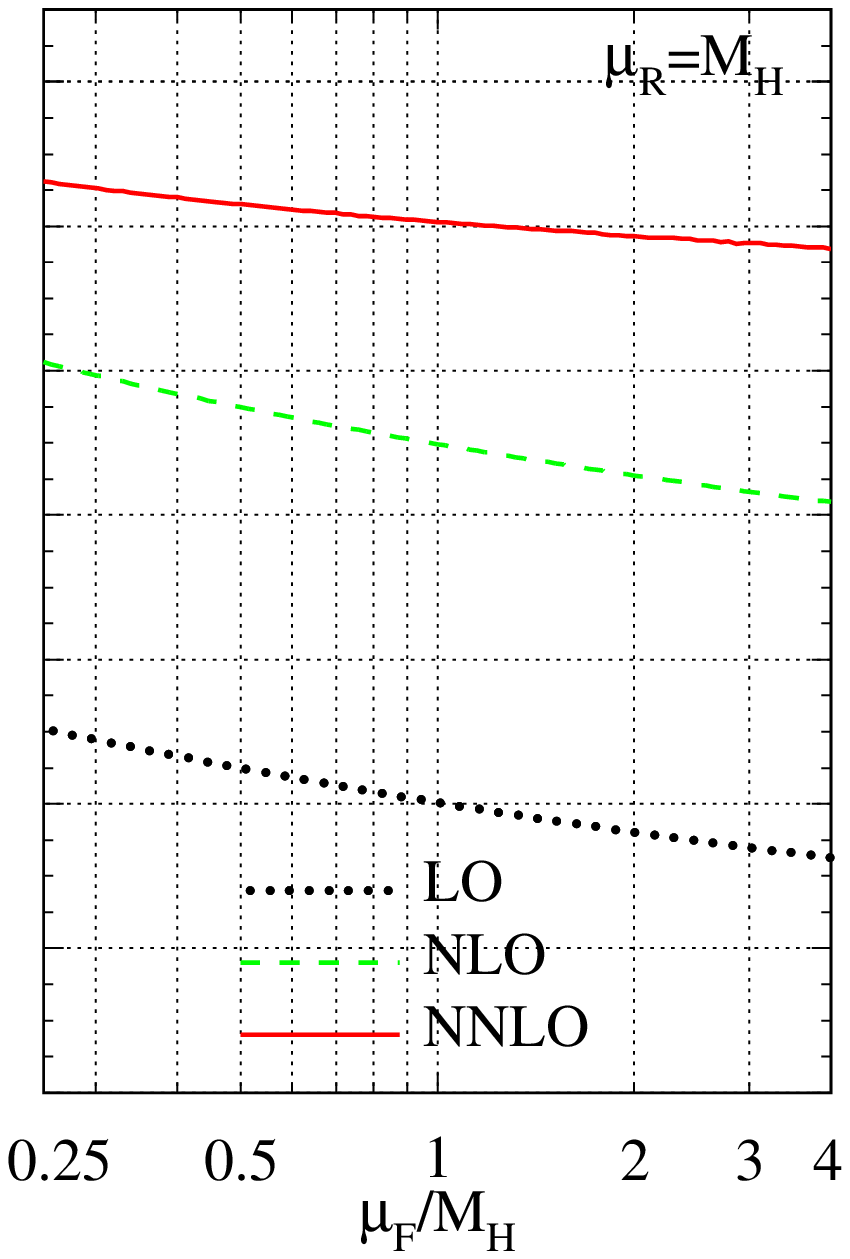} &
      \epsfxsize=12em
      \epsffile[184 210 411 610]{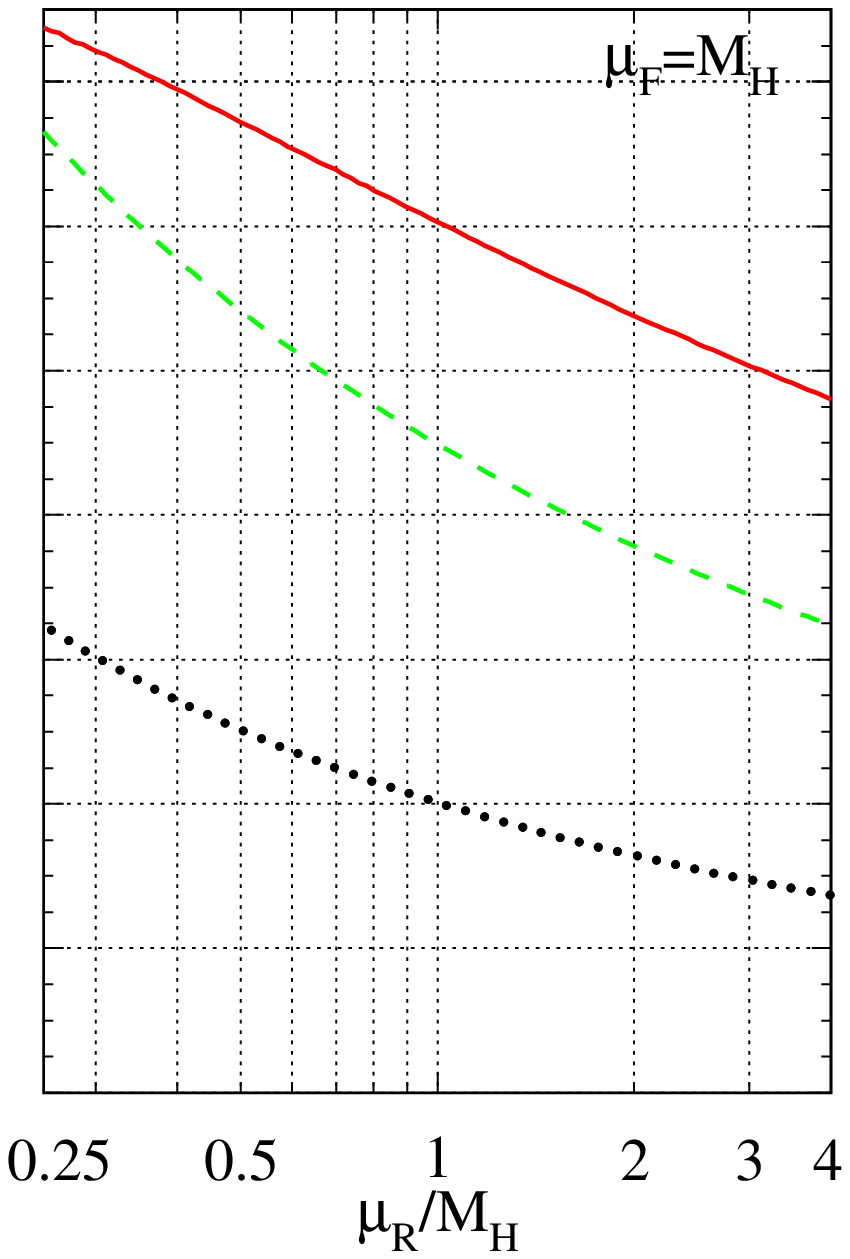}\\[-2em]
    \end{tabular}
      \caption[]{\label{fig::scale2}\sloppy
        Scale dependence of the cross section for $M_H=115$\,GeV at the
        Tevatron Run~II ($\sqrt{s} = 2$\,TeV).
        Left: variation of $\mu\equiv\mu_R=\mu_F$;
        center: variation of $\mu_F$ with $\mu_R=M_H$;
        right: variation of $\mu_R$ with $\mu_F=M_H$.
        }
  \end{center}
\end{figure*}

%- }}}
%- {{{ fig::scale14:

%
\begin{figure*}
  \begin{center}
    \leavevmode
    \begin{tabular}{ccc}
      \epsfxsize=12em
      \epsffile[184 210 411 610]{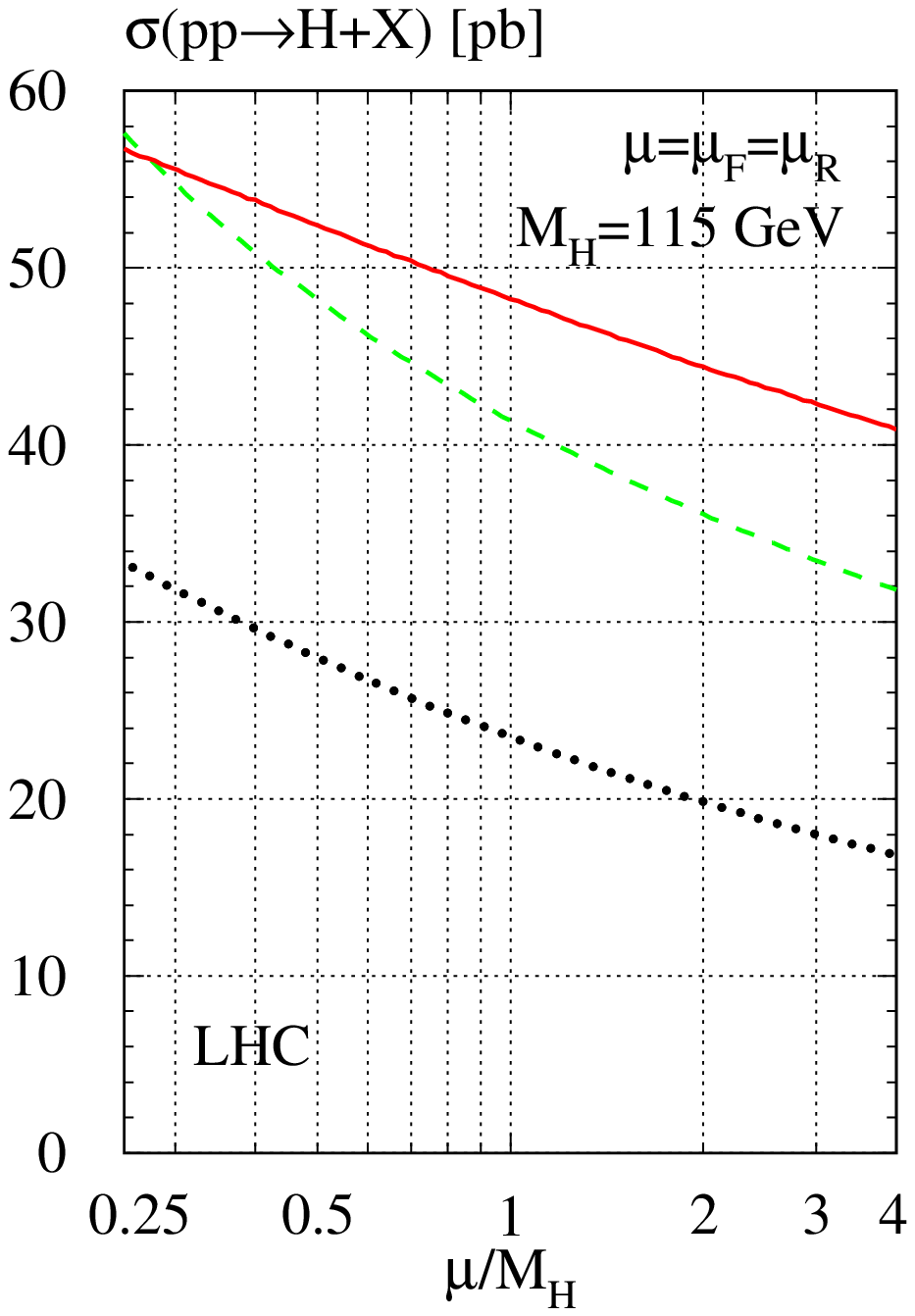} &
      \epsfxsize=12em
      \epsffile[184 210 411 610]{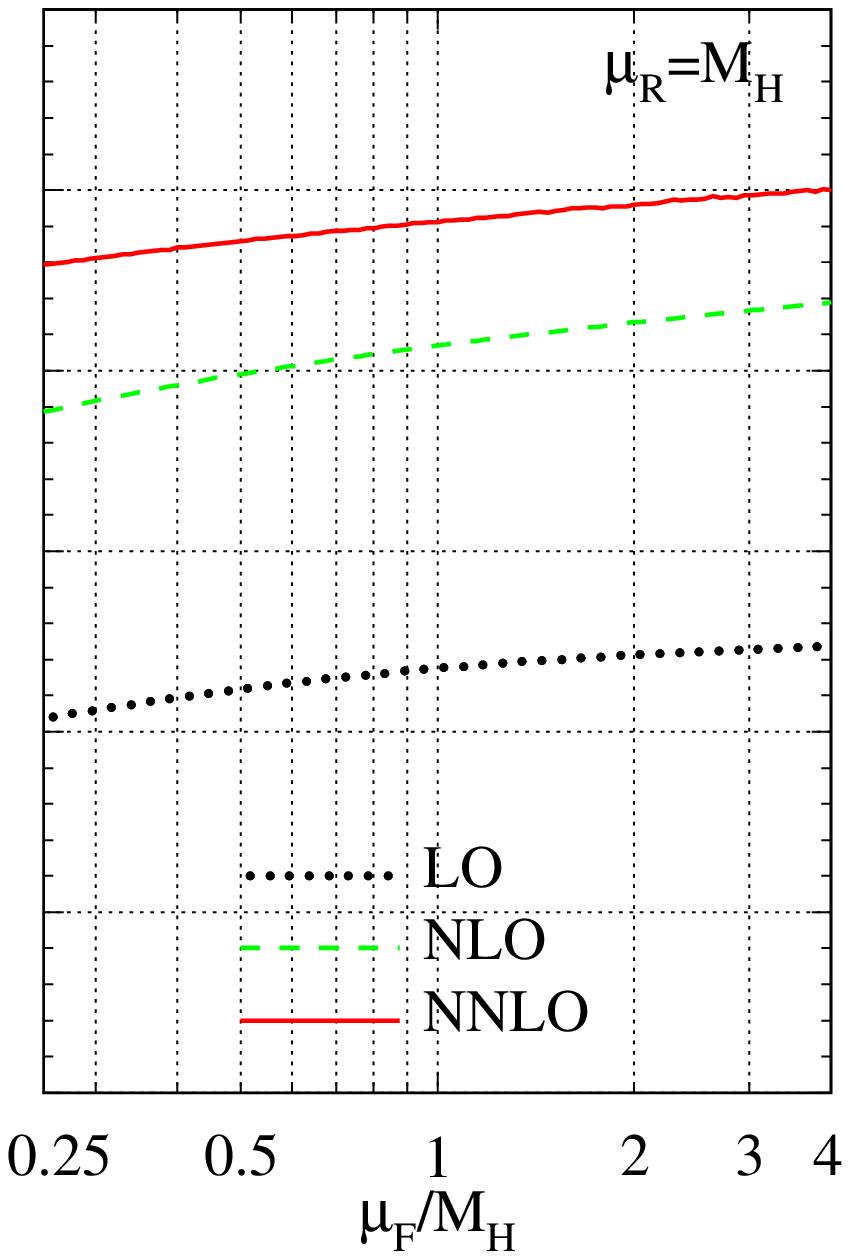} &
      \epsfxsize=12em
      \epsffile[184 210 411 610]{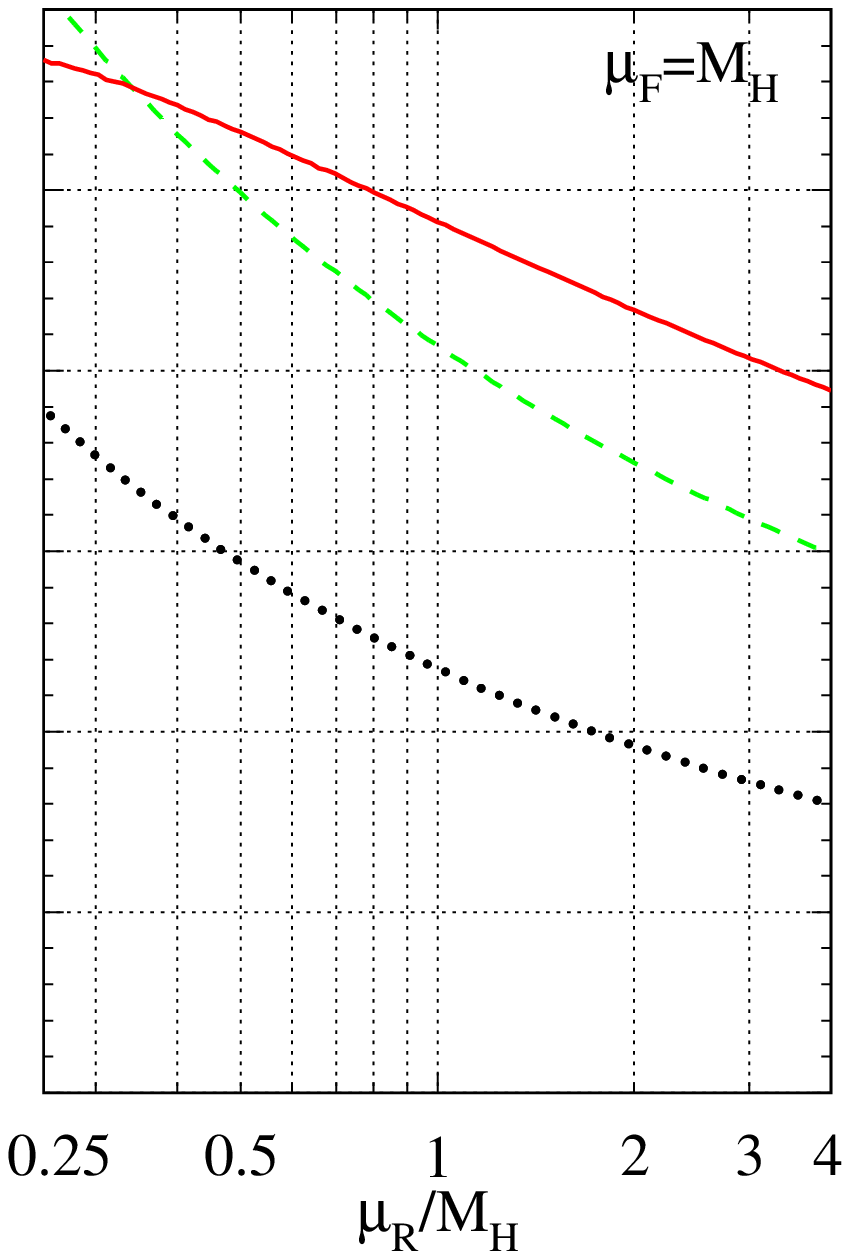}\\[-2em]
    \end{tabular}
      \caption[]{\label{fig::scale14}\sloppy
        Same as \fig{fig::scale2}, but for the {\abbrev LHC}.
        }
  \end{center}
\end{figure*}

%- }}}
%
Let us now present the results for the production of a Standard Model
Higgs boson at the Tevatron and the {\abbrev LHC}~\cite{HKggh,AnaMel}.
\fig{fig::snnlo} shows the total cross section for Higgs production at
$(a)$ the Tevatron, and $(b)$ the {\abbrev LHC}. One observes a nicely
converging perturbative series, together with a clear reduction of the
scale dependence in both cases.  Soft gluon resummation increases the
{\abbrev NNLO} curves by about 10\%, confirming the stability of the
perturbative result~\cite{resum}.

To investigate the scale dependence in more detail, we show the
variation of the cross section with respect to the renormalization and
the factorization scale $\mu_R$ and $\mu_F$ for a fixed Higgs mass of
$M_H=115$\,GeV at the Tevatron and the {\abbrev LHC} in \fig{fig::scale2}
and \ref{fig::scale14}, respectively. In the left panel of each figure,
$\mu_R$ and $\mu_F$ are identified and varied simultaneously within the
rather conservative range of $M_H/4 < \mu < 4M_H$. In the center panels,
the renormalization scale $\mu_R$ is identified with $M_H$, while
$\mu_F$ is varied, and in the right panels, $\mu_F$ is fixed, and
$\mu_R$ is varied. Using the variation between $M_H/2$ and $2M_H$ as an
indication of the theoretical uncertainty, one arrives at the conclusion
that the cross section at the Tevatron is known to about $\pm 15\%$, at
the {\abbrev LHC} to about $\pm 10\%$.

%- }}}
%- {{{ Pseudo-scalar Higgs production:

\section{Pseudo-scalar Higgs production}
The method described in \sct{sec::nnlo} can also be applied to evaluate
the production rate of a pseudo-scalar Higgs boson, as it is predicted
in extended models like the {\abbrev MSSM}. One should keep in mind,
however, that the couplings of the Higgs bosons may be altered in such
theories.  In the {\abbrev MSSM}, for example, the coupling to bottom
quarks might be enhanced by large values of $\tan\beta$. In this case,
bottom loop contributions cannot be neglected even at higher orders in
{\abbrev QCD}.

In \reference{ps}, the top loop contribution to the total rate for
pseudo-scalar Higgs production was evaluated up to \nnlo{}
{\abbrev QCD}. The corrections were found to be very similar to the
scalar Higgs case, while the leading order result determines the
different overall normalization. In consequence, within the limit where
bottom quark contributions can be neglected, the production rate for
pseudo-scalar Higgs bosons is known to the same level of accuracy as for
scalar Higgs bosons.

%- }}}
%- {{{ Conclusions:

%\section{Conclusions}
%
\vspace{1em} {\bf Conclusions.}\quad The production rate for scalar and
pseudo-scalar Higgs bosons has been shown to be described by a
well-behaved perturbative series up to {\abbrev NNLO}.  The newly
developed calculational methods of~\reference{HKggh} and
\reference{AnaMel} should prove useful also in many other applications.

%- }}}
%- {{{ bibliography:
%- {{{ journals:

\def\app#1#2#3{{\it Act. Phys. Pol. }\jref{\bf B #1}{#2}{#3}}
\def\apa#1#2#3{{\it Act. Phys. Austr. }\jref{\bf#1}{#2}{#3}}
\def\annphys#1#2#3{{\it Ann. Phys. }\jref{\bf #1}{#2}{#3}}
\def\cmp#1#2#3{{\it Comm. Math. Phys. }\jref{\bf #1}{#2}{#3}}
\def\cpc#1#2#3{{\it Comp. Phys. Commun. }\jref{\bf #1}{#2}{#3}}
\def\epjc#1#2#3{{\it Eur.\ Phys.\ J.\ }\jref{\bf C #1}{#2}{#3}}
\def\fortp#1#2#3{{\it Fortschr. Phys. }\jref{\bf#1}{#2}{#3}}
\def\ijmpc#1#2#3{{\it Int. J. Mod. Phys. }\jref{\bf C #1}{#2}{#3}}
\def\ijmpa#1#2#3{{\it Int. J. Mod. Phys. }\jref{\bf A #1}{#2}{#3}}
\def\jcp#1#2#3{{\it J. Comp. Phys. }\jref{\bf #1}{#2}{#3}}
\def\jetp#1#2#3{{\it JETP Lett. }\jref{\bf #1}{#2}{#3}}
\def\jhep#1#2#3{{\small\it JHEP }\jref{\bf #1}{#2}{#3}}
\def\mpl#1#2#3{{\it Mod. Phys. Lett. }\jref{\bf A #1}{#2}{#3}}
\def\nima#1#2#3{{\it Nucl. Inst. Meth. }\jref{\bf A #1}{#2}{#3}}
\def\npb#1#2#3{{\it Nucl. Phys. }\jref{\bf B #1}{#2}{#3}}
\def\nca#1#2#3{{\it Nuovo Cim. }\jref{\bf #1A}{#2}{#3}}
\def\plb#1#2#3{{\it Phys. Lett. }\jref{\bf B #1}{#2}{#3}}
\def\prc#1#2#3{{\it Phys. Reports }\jref{\bf #1}{#2}{#3}}
\def\prd#1#2#3{{\it Phys. Rev. }\jref{\bf D #1}{#2}{#3}}
\def\pR#1#2#3{{\it Phys. Rev. }\jref{\bf #1}{#2}{#3}}
\def\prl#1#2#3{{\it Phys. Rev. Lett. }\jref{\bf #1}{#2}{#3}}
\def\pr#1#2#3{{\it Phys. Reports }\jref{\bf #1}{#2}{#3}}
\def\ptp#1#2#3{{\it Prog. Theor. Phys. }\jref{\bf #1}{#2}{#3}}
\def\ppnp#1#2#3{{\it Prog. Part. Nucl. Phys. }\jref{\bf #1}{#2}{#3}}
\def\sovnp#1#2#3{{\it Sov. J. Nucl. Phys. }\jref{\bf #1}{#2}{#3}}
\def\sovus#1#2#3{{\it Sov. Phys. Usp. }\jref{\bf #1}{#2}{#3}}
\def\tmf#1#2#3{{\it Teor. Mat. Fiz. }\jref{\bf #1}{#2}{#3}}
\def\tmp#1#2#3{{\it Theor. Math. Phys. }\jref{\bf #1}{#2}{#3}}
\def\yadfiz#1#2#3{{\it Yad. Fiz. }\jref{\bf #1}{#2}{#3}}
\def\zpc#1#2#3{{\it Z. Phys. }\jref{\bf C #1}{#2}{#3}}
\def\ibid#1#2#3{{ibid. }\jref{\bf #1}{#2}{#3}}

\newcommand{\jref}[3]{{\bf #1} (#2) #3}

%- }}}

%- }}}


\begin{thebibliography}{99}
%1
\bibitem{CarHab} M. Carena, H.E. Haber, hep-ph/0208209.

%2
\bibitem{lo} H.~Georgi, S.~Glashow, M.~Machacek, D.V.~Nanopoulos,
        \prl{40}{1978}{692}.

%3
\bibitem{nlo} S.~Dawson, \npb{359}{1991}{283}; S.~Dawson, R.P.~Kauffman,
        \prd{49}{1994}{2298}; M.~Spira, A.~Djouadi, D.~Graudenz,
        P.M.~Zerwas, \npb{453}{1995}{17}; \plb{318}{1993}{347};
        D.~Graudenz, M.~Spira, P.M.~Zerwas, \prl{70}{1993}{1372};
        M.~Spira, \fortp{46}{1998}{203}.


%4
\bibitem{coef} K.G.~Chetyrkin, B.A.~Kniehl, M.~Steinhauser,
  \prl{79}{1997}{353}; \npb{510}{1998}{61}.

%5
\bibitem{DY} R.~Hamberg, T.~Matsuura and W.L.~van~Neerven,
  \npb{359}{1991}{343}; erratum \ibid{B 644}{2002}{403}.

%6
\bibitem{baismi} P.A. Baikov, V.A. Smirnov, \plb{477}{2000}{367}.

%7
\bibitem{form} J.A.M.~Vermaseren, math-ph/0010025;
         S.A. Larin, F.V. Tkachov, J.A.M. Vermaseren,
         NIKHEF-H/91-18, Amsterdam, 1991.


%8
\bibitem{qgraf} P.~Nogueira, \jcp{105}{1993}{279}.

%9
\bibitem{geficom} K.G.~Chetyrkin, M.~Steinhauser, unpublished.

%10
\bibitem{auto} R.~Harlander, M.~Steinhauser, \ppnp{43}{1999}{167}. 

%11
\bibitem{virtual} R.V. Harlander, \plb{492}{2000}{74}.
  
%12
\bibitem{kls} M.~Kr\"amer, E.~Laenen, M.~Spira, \npb{511}{1998}{523}.
  

%13
\bibitem{soft} R.V. Harlander, W.B. Kilgore, \prd{64}{2001}{013015}.

%14
\bibitem{cfg} S.~Catani, D.~de~Florian, M.~Grazzini, \jhep{0105}{2001}{025}.

%15
\bibitem{HKggh} R.V. Harlander, W.B. Kilgore, \prl{88}{2002}{201801}.

%16
\bibitem{kichep02} W.B. Kilgore, proceedings of {\small ICHEP'02},
  hep-ph/0208143.


  
%17
\bibitem{AnaMel} C. Anastasiou, K. Melnikov, \npb{646}{2002}{220}.

%18
\bibitem{AnaRADCOR} C.~Anastasiou, L.~Dixon, K.~Melnikov,
  hep-ph/0211141, these proceedings.

%19
\bibitem{mrstnnlo} A.D. Martin, R.G. Roberts, W.J. Stirling,
  R.S. Thorne, \plb{531}{2002}{216}.

%20
\bibitem{moments} W.L. van Neerven, A. Vogt, \plb{490}{2000}{111};
  A. Ret\'ey, J.A.M. Vermaseren, \npb{604}{2001}{281}.

%21
\bibitem{resum} S.~Catani, D.~de~Florian, M.~Grazzini, P.~Nason, {\it
    in:} W.~Giele {\it et al.}, ``The QCD/SM Working Group: Summary
    Report'', Les Houches 2001, hep-ph/0204316.

%22
\bibitem{ps} R.V.~Harlander, W.B.~Kilgore, \jhep{0210}{2002}{017};
        C.~Anastasiou, K.~Melnikov, hep-ph/0208115.


\end{thebibliography}
\end{document}